\documentclass{IAU}
\usepackage{graphicx} 
\usepackage{subfigure} 
\usepackage{wrapfig}
\usepackage{cleveref}

\title[Transitional disk archeology from exoplanet population synthesis] 
{Transitional disk archeology from exoplanet population synthesis}

\author[Germ\'an Chaparro Molano]   
{Germ\'an Chaparro Molano$^1$, Frank Bautista$^2$, Yamila Miguel$^3$
}

\affiliation{$^1$Vicerrector\'ia de Investigaci\'on, Universidad ECCI \\ Bogot\'a, Colombia \\ email: {\tt gchaparrom@ecci.edu.co} \\[\affilskip]

$^2$Departamento de F\'isica, Universidad Nacional de Colombia \\ Bogot\'a, Colombia
\\ email: {\tt fjbautistas@unal.edu.co} \\[\affilskip]
$^3$Sterrewacht Leiden, Leiden University \\ Leiden, The Netherlands
}

\pubyear{2019}
\volume{345}  
\jname{Origins: from the Protosun to the First Steps of Life}
\editors{Bruce G. Elmegreen, L. Viktor T\'oth, Manuel G\"udel, eds.}

\begin{document}

\maketitle

\begin{abstract}
Increasingly better observations of resolved protoplanetary disks show a wide range of conditions in which planets can be formed. Many transitional disks show gaps in their radial density structure, which are usually interpreted as signatures of planets. It has also been suggested that observed inhomogeneities in transitional disks are indicative of dust traps which may help the process of planet formation. However, it is yet to be seen if the configuration of fully evolved exoplanetary systems can yield information about the later stages of their primordial disks. We use synthetic exoplanet population data from Monte Carlo simulations of systems forming under different density perturbation conditions, which are based on current observations of transitional disks. The simulations use a core instability, oligarchic growth, dust trap analytical model that has been benchmarked against exoplanetary populations.
\end{abstract}
\firstsection 
\section{Introduction}

Planet-forming disks can have either smooth or density perturbed profiles
(\cite[van der Marel et al. 2015]{Marel_etal15}). Gas-depleted, density perturbed
disks are often called transitional disks. Gaps, cavities, or radial
perturbations in such disks sometimes show dust traps in which planet formation
may be taking place. Such dust traps may be caused by pressure bumps, which are
often theorized as a consequence of the presence of an already-formed planet in
the disk (\cite[Pinilla et al. 2011]{Pinilla_etal11}). In this work, we are
interested in the impact of disk density perturbations in planetary populations
while making no assumptions about the exact mechanism that causes such
inhomogeneities in the disk. In order to explore the link between transitional
disks and exoplanetary systems, we intend to compare synthetic and observed
populations of exoplanetary systems. However, instead of looking at individual
cases (\cite[Raymond et al. 2018]{Raymond_2018}) we take a Bayesian inference
approach to reconstruct probability distributions of general properties of 3000+
simulated synthetic planetary systems. We thus study the effect of radial density
perturbations in the disk structure on the formation of exoplanetary systems. For
many transitional disks, a radial density perturbation can be described as a
succession of over-dense and under-dense regions which appear as the radial
distance $r$ changes (\cite[Pinilla et al. 2011]{Pinilla_etal11}),
\begin{equation}\label{ec:sigmap}
\Sigma_p(r)=\Sigma(r)\left(1+A\cos\left(2\pi\frac{r}{fH(r)}\right)\right)\ .
\end{equation}
Here $\Sigma(r)$ is the surface density distribution (\cite[Miguel et al.
2011]{Miguel_etal11}), $A$ is the amplitude of the perturbation, $f$ is the
length scale of the perturbation and $H(r)$ is the scale height of the disk. Here
we consider $A=0$ for smooth disks, and $A=0.3$ for transitional disks.

\section{Synthetic planetary systems}

Population synthesis models are often used for modeling individual exoplanetary
systems (\cite[Raymond et al. 2018]{Raymond_2018}) in order to estimate
properties of individual planets in the system. We extend this method for 3000
synthetic planetary systems formed  in smooth and transitional disks (following
the perturbation recipe described above). Each system has initial conditions
drawn from prior probability distributions on stellar mass, disk mass and radial
extent, stability, metallicity, and gas dissipation timescale from \cite[Miguel
et al. (2011)]{Miguel_etal11}. This planet population synthesis framework is also
described as part of the review in \cite[Benz et al. (2014)]{Benz_etal14}. The
result of each simulation is a system with orbital data like planetary semi-major
axis and mass (solid+gas). We calculate consolidated quantities that summarize
general properties of each simulated system. Thus, we use quantities like the
number of terrestrial and giant planets, total terrestrial planetary mass,
average terrestrial planet mass, total planetary mass/disk mass ratio, and what
we refer here to as center of mass. This center of mass is not the barycenter of
the system but rather the first moment of the mass distribution of planets with
respect to their semi-major axes. With the consolidated results per system, we
can reconstruct the posterior probability distribution from the simulated
quantities mentioned above. This reconstructed posterior can be used to make
educated predictions for existing exoplanetary systems based on known properties
like stellar mass, metallicity, planetary masses and semi-major axes
distributions, etc.



\section{Results}

\begin{figure}[h]
\centering
\subfigure[]{\includegraphics[width=0.45\textwidth]{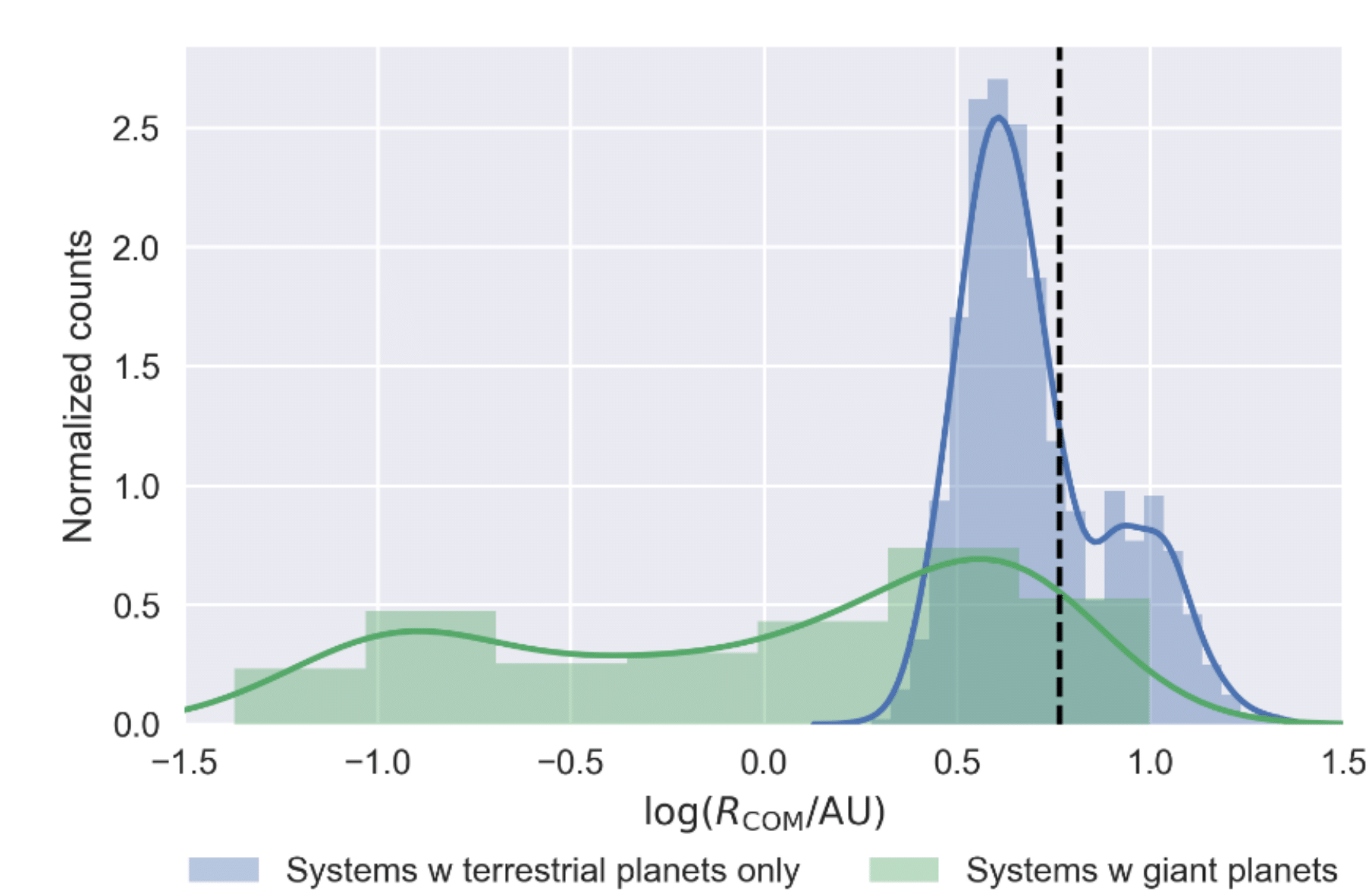}\label{fig:check1}}
\subfigure[]{\includegraphics[width=0.43\textwidth]{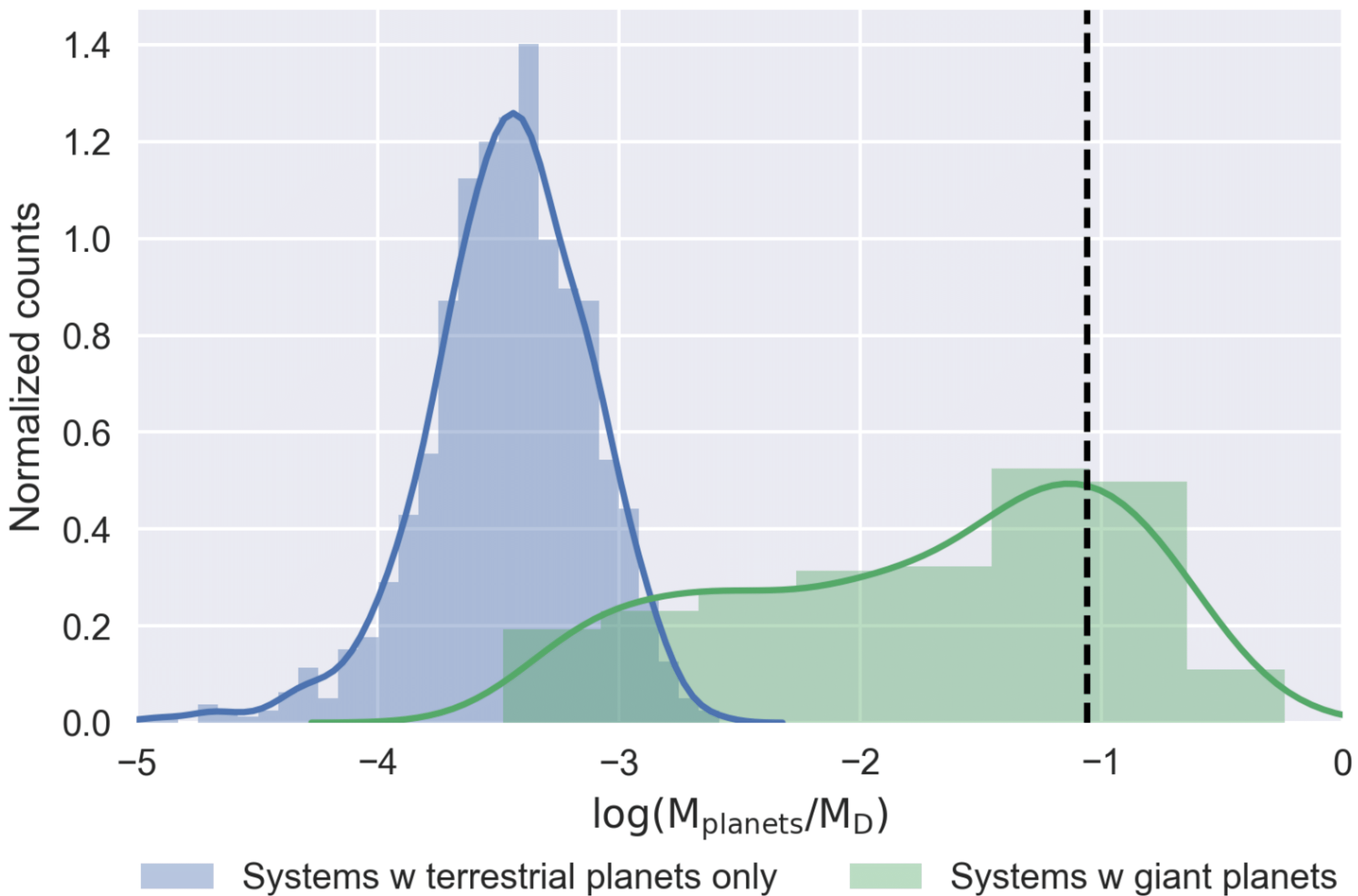}\label{fig:check2}}
\caption{Marginalized distributions of center of mass and total planet mass/disk mass ratio for systems with giant planets (green) and with only terrestrial planets (blue). The distributions are approximated by a Kernel Density Estimation (solid lines).} \label{fig:check}
\end{figure}

\begin{figure}[h]
\begin{center}
 \includegraphics[width=0.6\textwidth]{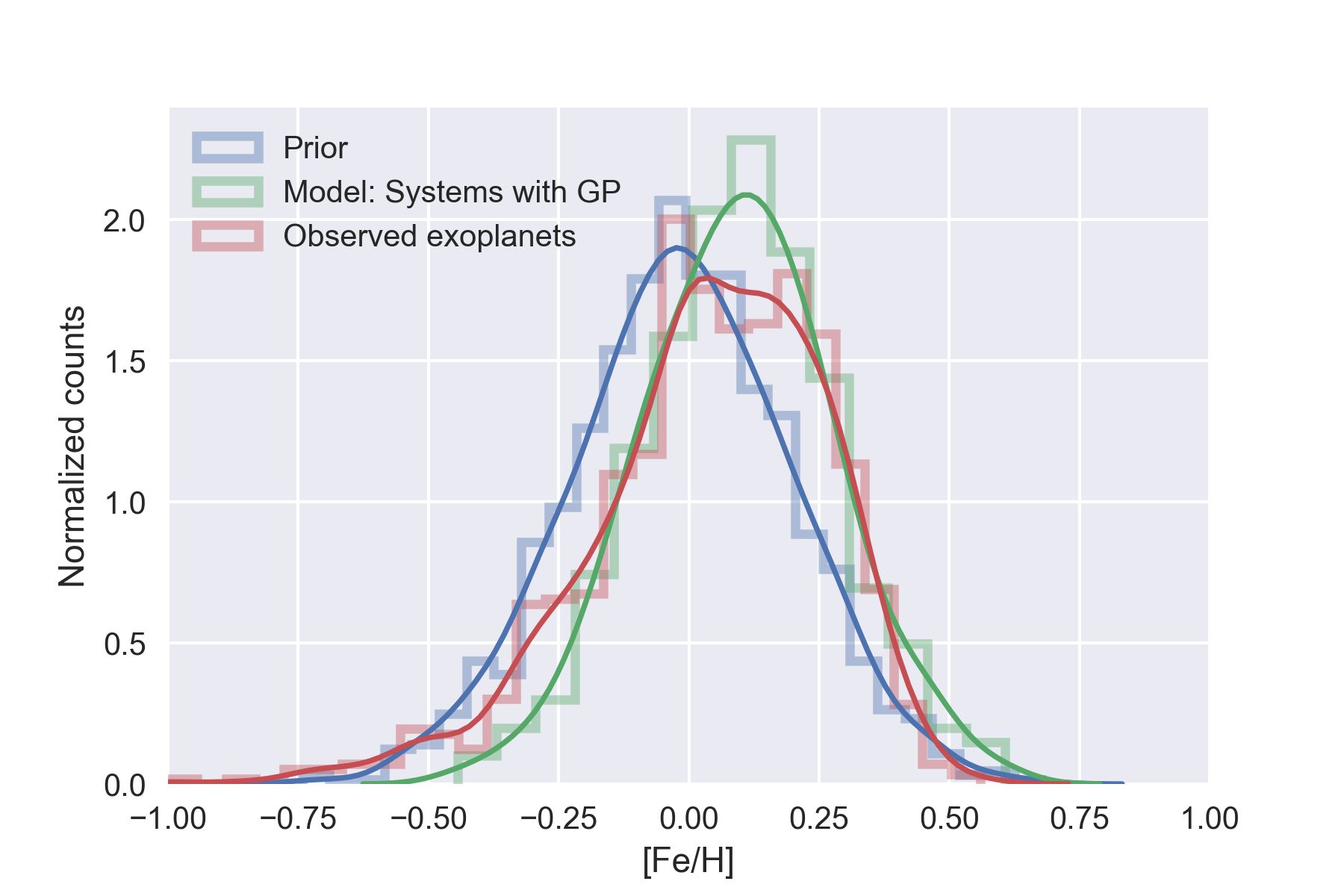}
 \caption{Marginalized distribution of exoplanetary systems metallicities for synthetic
 systems with giant planets (green), observed systems with giant planets (red) and from
 the prior used in the simulation (blue) from \cite[Mayor et al. (2011)]{Mayor_etal11}. The distributions are
  approximated by a Kernel Density Estimation (solid lines).}
   \label{fig:Metalhist}
\end{center}
\end{figure}

\begin{figure}[h]
\begin{center}
 \includegraphics[width=0.7\textwidth]{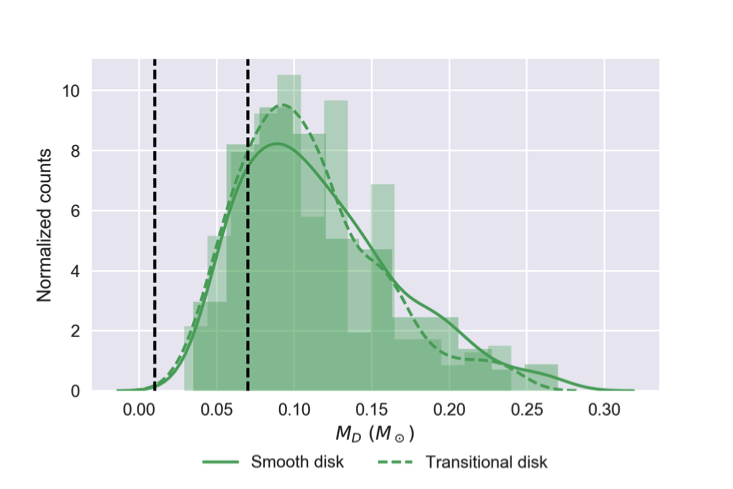}
 \caption{Marginalized posterior distribution of parent disk masses for synthetic systems with giant planets for smooth and transitional disks. The dashed black lines show lower and upper estimations for the minimum mass solar nebula. The distributions are approximated by a Kernel Density Estimation (solid lines).}
   \label{fig:SvsT_1}
\end{center}
\end{figure}

\begin{figure}[h]
\begin{center}
    \includegraphics[width=0.8\textwidth]{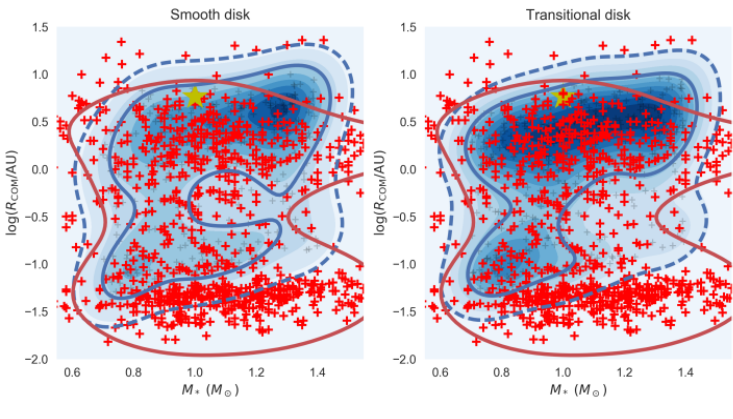}
    \caption{Center of mass vs. stellar mass for observed systems (red) and synthetic systems (blue) formed in smooth disks (left) and in transitional disks (right). The 1$\sigma$ (solid line) amd 2$\sigma$ (dashed line) contours were obtained using a Kernel Density Estimation for each set of data points.}
    \label{fig:NGP}
\end{center}
\end{figure}

As an initial benchmark, we looked at the resulting distributions of the location
of the center of mass and the total planetary mass/disk mass ratio for systems
formed in smooth disks. Figure \ref{fig:check1} shows that systems with giant
planets are more spread inwards (closer to the star) than systems that only
formed terrestrial planets, due to giant planet migration. On the other hand,
more of the original disk mass goes to form planets in systems with giant planets
than in terrestrial planet-only systems (Figure \ref{fig:check2}). Both of these
results are expected from exoplanet and planet-forming disk observations. The
resulting synthetic planetary systems were divided among systems with giant
planets (20-25\% of all systems), and systems with terrestrial planets only. We
benchmarked our results of the distribution of planetary systems with respect to
metallicity. Figure \ref{fig:Metalhist}), shows that systems with giant planets
tend to be formed in metal-rich systems, both in our simulations and in
observations from \texttt{exoplanet.eu}. Comparatively, most exoplanetary systems
discovered to date tend to be distributed uniformly around the solar value in the
HARPS 2011 catalog (\cite[Mayor et al. 2011]{Mayor_etal11}), which we used as a
prior for our simulations. For the same initial disk mass, transitional disks
form more systems with giant planets than smooth disks (Figure \ref{fig:SvsT_1}).
This is due to transitional disks having over-dense regions that favor planet
growth. Figure \ref{fig:NGP} shows the distribution of stellar mass vs. center of
mass for synthetic and observed planetary systems. The 1$\sigma$ and 2$\sigma$
contours show that there is a significant overlap in parameter space between
synthetic and observed planetary systems. For exoplanetary systems with a low
center of mass the overlap breaks down, which is likely a result of observational
bias.
\section{Conclusions}
Our comparison between smooth and transitional disks shows that transitional disks favor the formation of giant planets at lower disk masses than smooth disks. Benchmarking of the results of our simulations show a very good parameter space overlap between synthetic and observed exoplanetary systems. The method of approximating a posterior probability distribution for exoplanetary parameters from our simulations can be used to make educated predictions that direct future surveys of observed exoplanetary systems.
\begin{discussion}

\discuss{Alfaro} {What kind of priors are you using in the Bayesian inference 
methods?}

\discuss{Chaparro} {The functional form of priors for each input parameter are taken from the literature, which are summarized in the 2011 work by Miguel et al.}

\end{discussion}

\end{document}